\documentclass[aip,jcp,,preprint,letterpaper]{revtex4}
\usepackage{graphicx}
\usepackage{comment}
\usepackage{multirow}
\usepackage{array}
\usepackage{bm}
\usepackage{amsmath}
\usepackage{color}
\usepackage{ textcomp }
\usepackage{mathtools}
\begin{document}
\newcommand{\wn}{cm$^{-1}$}
\newcommand{\td}{$\sim$}
\newcommand{\la}{\langle}
\newcommand{\ra}{\rangle}
\newcommand{\e}{\epsilon}
\newcommand{\w}{\omega}
\newcommand{\bracket}[1]{\left\langle #1 \right\rangle}
\newcommand{\degreec}{^{\circ}{\rm C}}
\newcommand{\be}{\begin{equation}}
\newcommand{\ee}{\end{equation}}
\newcommand{\ie}{{\it i.e.}}
\newcommand{\eg}{{\it e.g.}}
\newcommand{\etal}{{\it et al.}}
\newcommand{\bra}[1]{\left<#1\right|}
\newcommand{\ket}[1]{\left|#1\right>}
\newcommand{\ketbra}[2]{\ket{#1}\bra{#2}}

\title{The Enhancement of Interfacial Exciton Dissociation by Energetic Disorder is a Nonequilibrium Effect}
\author{Liang Shi}\thanks{Contributed equally.}
\affiliation{Department of Chemistry, Massachusetts
Institute of Technology, Cambridge, Massachusetts 02139, USA}
\author{Chee Kong Lee}\thanks{Contributed equally.}
\affiliation{Department of Chemistry, Massachusetts
Institute of Technology, Cambridge, Massachusetts 02139, USA}
\author{Adam P. Willard} \email{awillard@mit.edu}
\affiliation{Department of Chemistry, Massachusetts
Institute of Technology, Cambridge, Massachusetts 02139, USA}

\date{\today}

\begin{abstract}
The dissociation of excited electron-hole pairs is a microscopic process that is fundamental to the performance of photovoltaic systems. For this process to be successful, the oppositely charged electron and hole must overcome an electrostatic binding energy before they undergo ground state recombination. Here we use a simple model of charge dynamics to investigate the role of molecular disorder in this process. 
This model reveals that moderate spatial variations in electronic energy levels, such as those that arise in disordered molecular systems, can actually increase charge dissociation yields. We demonstrate that this is a nonequilibrium effect that is mediated by the dissipation driven formation of partially dissociated intermediate states that are long-lived because they cannot easily recombine. We present a kinetic model that incorporates these states and show that it is capable of reproducing similar behavior when it is parameterized with nonequilibrium rates.
\end{abstract}
\maketitle

\clearpage
The dissociation of Coulombically bound excited electron-hole pairs -- excitons --  into free charge carriers is a microscopic process that is fundamental to the performance of photovoltaic systems.\cite{Zhu2015,Deibel2010,Clarke2010}
This process requires the physical separation of oppositely charged electrons and holes, which are initially held together by an attractive electrostatic force.
The energy required to overcome this force and produce independent charge carriers is known as the exciton binding energy. 
For inorganic-based photovoltaic materials, the binding energy is generally small and easily overcome, however, for organic-based photovoltaics (OPVs) the exciton binding energy can significantly exceed thermal energies.
The inability of bound charges to overcome this large binding energy has been implicated as a primary source of efficiency loss in OPVs.\cite{Zhu2015,Deibel2010,Clarke2010,Bredas2009,vandewal2009,Veldman2009}
Many efforts to improve OPV efficiency have thus aimed to extend exciton lifetimes and enhance charge carrier mobilities by eliminating sources of microscopic disorder within the active material.\cite{Hains2010,Campoy-Quiles2008,Moule2008}
In this manuscript we reveal that this general strategy can have unintended negative effects on photovoltaic efficiency.
In particular, we demonstrate that the presence of molecular disorder can actually enhance exciton dissociation yields by giving rise to dissociation pathways that are downhill in energy and thus mitigate the effects of the exciton binding energy.
Using a simple model of exciton dynamics we show that when disorder is present electrons and holes are driven apart along these energetically favorable pathways.
We highlight that this effect is driven by the dissipation of excess electronic energy and is therefore determined by the nonequilibrium dynamics of the electron-hole pair.
Our results provide new physical insight into the importance of treating nonequilibrium effects in models of charge and energy transport.

In OPV materials exciton dissociation is facilitated by donor-acceptor interfaces, where energetic offsets in the molecular orbital energies of donor and acceptor molecules provide a driving force for exciton dissociation.
This driving force favors the formation of partially dissociated charge-transfer (CT) states, where the electron and hole reside on adjacent acceptor and donor molecules respectively.
These \textit{bound} CT states are further stabilized by the electrostatic attraction of the oppositely charged electron and hole, which is typically about 0.4 eV ($\sim10k_\mathrm{B}T$ at room temperature), and this strong Coulombic stabilization causes the bound CT state to lie at a minimum of the excited state potential energy surface.\cite{Zhu2009,Monahan2015,Muntwiler2008}
CT states that reside within this minimum are prone to recombination on timescales that are much shorter those required for the electron-hole pair to diffusively overcome the exciton binding energy.
Based on the Onsager model\cite{Onsager1938} the dissociation probability for a bound CT state at a typical organic donor-acceptor interface is approximately $P_\mathrm{dis} \sim 10^{-3}$.\cite{footnote1} 
Despite this exceedingly small prediction, the highest performing organic solar cells have been observed to operate with the internal quantum efficiency of near 100\%,\cite{Park2009} indicating that free change carriers escape this minimum with near unit efficiency.
Reconciling the apparent inconsistency between the predicted and observed recombination losses has been a longstanding challenge in the field of organic electronics.

Many studies, both experimental and theoretical, have been aimed at investigating how electron-hole pairs escape, or otherwise avoid, the trap-like bound CT state.\cite{Barth1997,Rubel2008,Lee2010,Gregg2011,Bakulin2012,VanEersel2012,
Yost2013,Grancini2013,Jailaubekov2013,Gelinas2014,Tscheuschner2015,Kocherzhenko2015,
Monahan2015,Hood2016,Stolterfoht2016,DAvino2016,Athanasopoulos2017}
Numerous plausible explanations have emerged from these efforts.
Experiments show that for some systems successful dissociation pathways avoid the lowest energy CT intermediates by traversing a non-thermalized manifold of high-energy, often delocalized, electronic states.\cite{Grancini2013,Jailaubekov2013} 
In other systems it has been shown that these so-called hot CT states are not necessary for dissociation and that free carriers can emerge from populations of electronically thermalized low-energy CT states.\cite{Lee2010,Gelinas2014,Tscheuschner2015}
The microscopic mechanism underlying this \textit{cold} exciton dissociation process remains a topic of scientific debate and is the focus of the work presented here.

Previous studies have identified various physical driving forces that may contribute favorably to the process of cold exciton dissociation.
This includes those arising from entropic effects, the presence of static interfacial electric fields, interfacial gradients in molecular excitation energies, and delocalized free carrier wave functions.\cite{Lee2010,Gregg2011,VanEersel2012,Yost2013,Gelinas2014,Monahan2015,Kocherzhenko2015,
Tscheuschner2015,Hood2016,DAvino2016,Athanasopoulos2017}
These contributions, and others, are generally sensitive to the presence of random molecular disorder, which can affect the inter- and intra-molecular electronic structure, leading to spatial variations in the energetic properties of excitons and free charge carriers.
Such disorder is common in organic electronic materials, however, its effect on the microscopic dynamics of electrons and holes is yet to be fully appreciated. 
Recently it has been found that the presence of random energetic disorder can both reduce the free energy barrier and enhance the thermodynamic driving force for exciton dissociation.\cite{Gregg2011,Govatski2015,Hood2016}
Here we expand upon this finding by exploring the effect of random energetic disorder on the microscopic dynamics of exciton dissociation.
By doing so we reveal that when disorder is present dissociation occurs primarily along nonequilibrium pathways and thus it cannot be properly understood in terms of thermodynamics alone.

Spatial variations in the energy landscape influence the dynamics of excitons and free charge carriers by biasing their motion along energetic gradients toward regions that permit the population of lower energy excited states.
In time-resolved fluorescence microscopy this effect manifests as a concerted red-shift and spatial broadening of the photoluminescence profile.\cite{Akselrod2014,Deotare2015}
In some cases spatial energetic variations can stabilize states that would be unfavorable within a perfectly ordered system.
For instance, at a donor-acceptor interface this effect can stabilize the formation of CT states with increased electron-hole separation, thereby facilitating the nascent stages of exciton dissociation.
As we demonstrate, this stabilization has a positive effect on the dissociation process that increases with disorder but also competes with a concomitant decrease in charge carrier mobility.
These competing effects combine to predict exciton dissociation yields that are maximized with a moderate amount of molecular disorder.

In the following section we describe the details of our model system. 
Then, in Sections~\ref{sec:dis} through ~\ref{sec:noneq} we present the results of our investigation, highlighting disorder's influence on both the equilibrium (\textit{i.e.}, thermodynamic) and nonequilibrium driving forces.
In Sec.~\ref{sec:kin} we demonstrate that the nonequilibrium effects of disorder on dissociation dynamics can be captured in the context of a simple kinetic model.
Finally, in Sec.~\ref{sec:imp}, we discuss the implications of our finding for modern organic electronics.

\section{A Coarse-Grained Model of Charge-Transfer Exciton Dynamics}
To simulate the effect of nanoscale disorder on exciton dissociation requires system sizes and time scales that are well beyond the capability of modern quantum chemistry.
Fortunately, recent work has revealed that dynamics of CT excitations can be accurately described using a simple and efficient coarse-grained model of incoherent charge carrier dynamics.\cite{Deotare2015}
Our investigation utilizes this theoretical framework to reveal the fundamental relationship between static molecular disorder and the dissociation of CT excitons.
Our model does not include any specific atomistic-level detail, nor does it include high-level information about the electronic structure.
Nonetheless, as we have previously demonstrated,\cite{Deotare2015,Lee2016a} when this model is parameterized appropriately it exhibits the remarkable ability to simultaneously reproduce multiple experimental observations related to the dynamics of CT excitons.

As illustrated in Fig.~\ref{fig:1}(a), our model describes the system as a collection of individual molecules arranged on a square lattice and separated into a donor phase and an acceptor phase.
We describe the presence of molecular disorder by assigning each molecule a HOMO or LUMO energy, denoted $\epsilon_\mathrm{HOMO}$ or $\epsilon_\mathrm{LUMO}$ respectively, drawn randomly from a Gaussian distribution, $P(\epsilon)=(2\pi \sigma^2)^{-1/2}\exp[-(\epsilon-\epsilon^{(0)})^2/2\sigma^2]$.
Here $\epsilon^{(0)}$ denotes the average orbital energy and $\sigma$ defines the width of the site energetic distribution.
We control the amount of disorder within the system by varying the width of the Gaussian distribution, indicated in terms of $\sigma$.
Charge transfer excitations are modeled as point particles of opposite charge (\textit{i.e.}, electron and hole) that are localized on separate donor and acceptor molecules.
The potential energy of a given CT state is given by $E=E^{(\mathrm{Coul})}+E^{(\mathrm{vert})}$, where $E^{(\mathrm{Coul})}$ denotes the electrostatic interaction of the electron and hole and $E^{(\mathrm{vert})}$ is the HOMO-LUMO gap of the  specific donor-acceptor pair that is occupied, (i.e., $E^{(\mathrm{vert})}$ is given by the difference between the values of $\epsilon_{\mathrm{LUMO}}$ of the electron's site and $\epsilon_{\mathrm{HOMO}}$ of the hole's site).
The time evolution of the CT state is determined by a kinetic Monte Carlo (KMC) algorithm that simulates the asynchronous hopping of electrons and holes.
The KMC algorithm also includes a ground state recombination process, which can only occur if the electron and hole reside on adjacent molecules.
This recombination process results in the termination of the trajectory.
The charge recombination rate and the charge hopping rates can be determined based on a combination of experimental inputs and theoretical models, as described within the Methods section.
We assume excitons are fully dissociated when the electron and hole are separated by a distance that is greater than the Coulomb radius, which we define as the distance for which $E^{(\mathrm{Coul})}=k_\mathrm{B}T$. 
Since this assumption neglects many contributions to trapping and recombination, our computed dissociation yields represent an upper limit of the actual process.
Notably, however, if such loss mechanisms are  independent of disorder, then the relative dissociation yields we compute are expected to be more accurate than their absolute values.

In the results presented below we have utilized the same model parameterization as in our previous studies.\cite{Deotare2015}
Although this parameterization has been optimized to describe the dynamics of a specific donor-acceptor blend, we take it to be representative of a generic small-molecule organic heterojunction.
We generate trajectories by randomly initializing electrons and holes on adjacent molecules along the donor-acceptor interface.
We do this to mimic the process of photoexcitation, which we assume yields a population distributed uniformly within the energetic density of states.
We neglect contributions from electronically hot CT states, meaning that CT states in our model are uniquely specified by the electron and hole positions.
For a given value of $\sigma$ we generate ensembles of trajectories by sampling the dynamics of many CT states over many different realizations of the random energetic disorder.
The analysis and interpretation of these trajectories are presented in the sections below.

\section{Dependence of Dissociation Yield on Disorder\label{sec:dis}}

\begin{figure}[h!tbp]
\includegraphics[width=8cm]{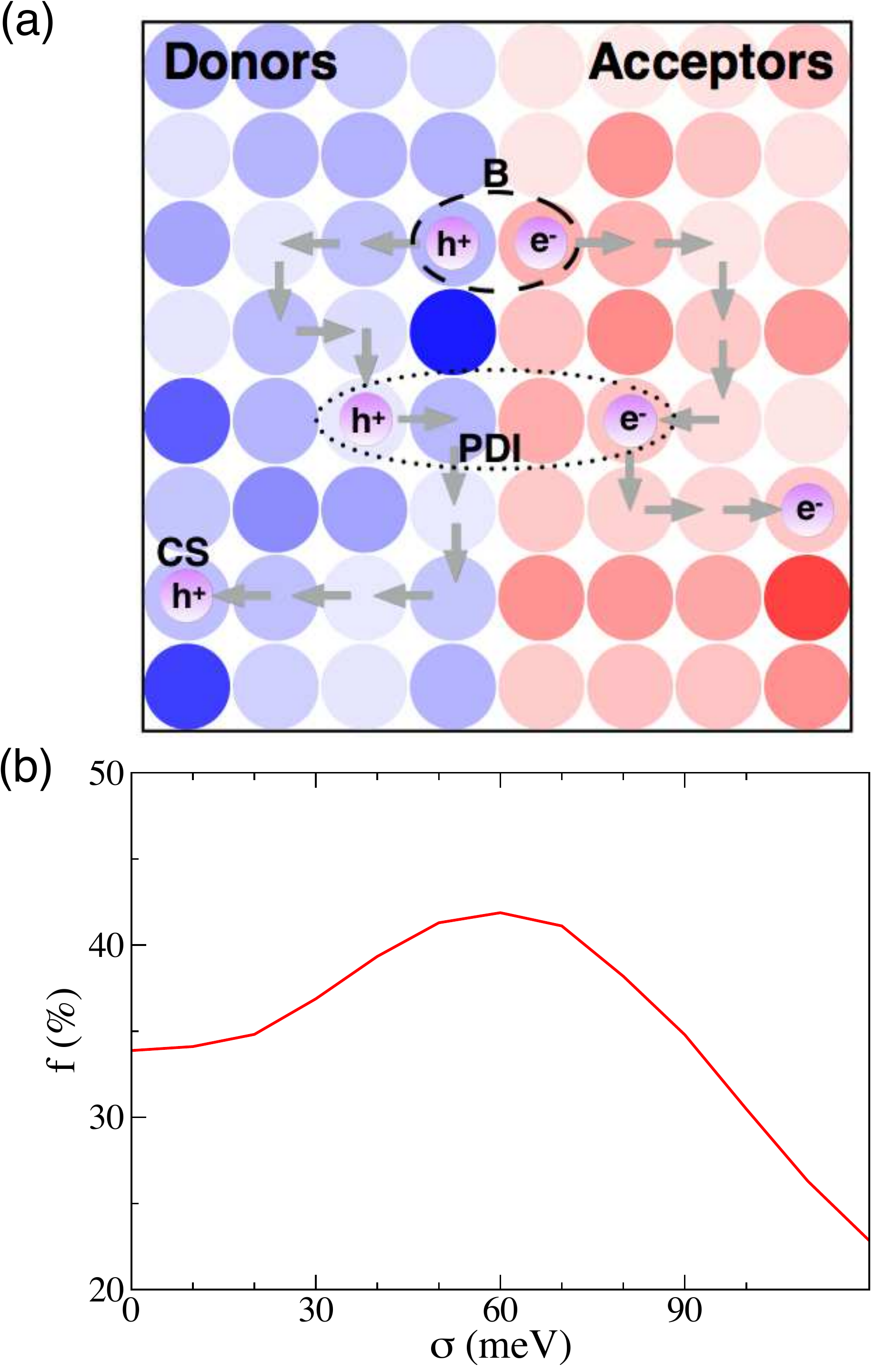}
\caption{
(a) A schematic of our model for simulating the dynamics of interfacial CT excitons. 
The color shadings of the blue and red circles represent the varying HOMO energies of donor molecules and LUMO energies of acceptor molecules, respectively. 
The circles with h$^+$ and e$^-$ are the hole and electron in the CT exciton, respectively. 
A representative trajectory is shown as grey arrows, where the electron and hole break apart gradually from the bound CT state (B), to the partially dissociated intermediate state (PDI), and finally to the fully dissociated state (CS). The definitions of the B, PDI, and CS states are given in the main text. 
(b) The dependence of the CT exciton dissociation yield, $f$, on the energetic disorder, $\sigma$. 
}
\label{fig:1}
\end{figure}

To study the effect of disorder on exciton dissociation we analyze ensembles of trajectories generated at various values of $\sigma$.
For a given value of $\sigma$ we determine the dissociation yield, $f$, by computing the fraction of trajectories that avoid recombination and escape the Coulomb capture radius (about 16 nm in this system). 
The plot in Fig.~\ref{fig:1}b illustrates that the dissociation yield depends non-monotonically on the amount of disorder in the system.
In other words, exciton dissociation is maximized in systems that include a finite amount of energetic disorder.

This finding is not without precedent.
Disorder-induced increases in exciton dissociation yields have been demonstrated previously in experiment \cite{Barth1997} and in simulation studies.\cite{Albrecht1995,Offermans2005,Govatski2015,Hood2016}
At the same time, it is well known that high levels of disorder lead to efficiency loss due to reduction in charge transport properties.\cite{Bassler1993}
Taken together, these competing effects suggest the existence of a maximum in $f$ at some optimal level of energetic disorder.
Despite this, the microscopic origins of these effects, and how their interplay mediates exciton dissociation, remains uncharacterized.
Our model study addresses this problem by identifying the nonequilibrium effects that are responsible for a disorder-induced enhancement in the CT dissociation process.

\section{The Effect of Disorder on the Thermodynamics of Exciton Dissociation\label{sec:eq}}
\begin{figure}[h!tbp]
\includegraphics[width=5.cm]{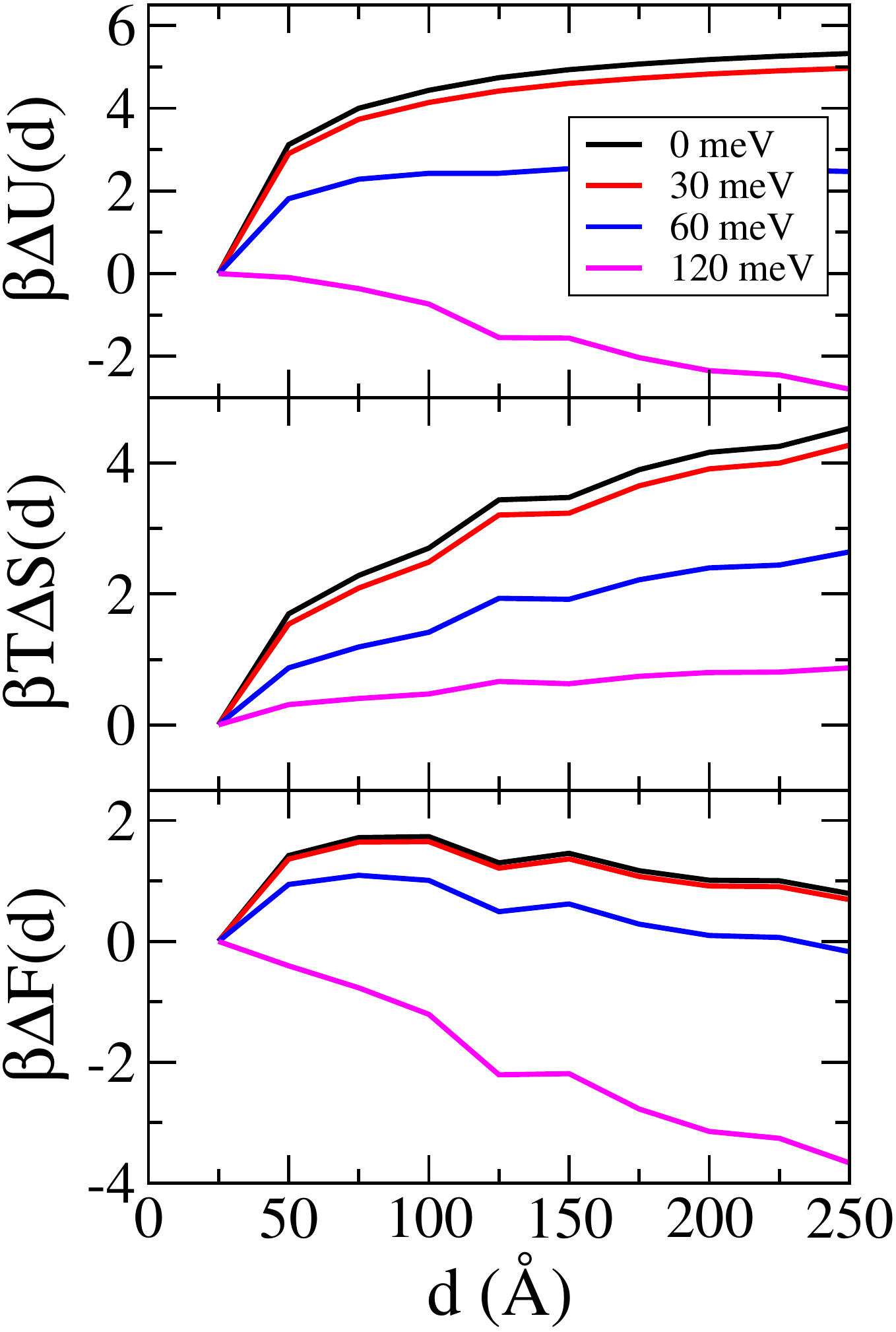}
\caption{
The internal energy (top), entropy (middle), and Helmholtz free energy (bottom) profiles as functions of electron-hole separation, 
$d$, as evaluated using Eq. (\ref{eq:Fofd}) - (\ref{eq:ent}). 
All the thermodynamic quantities are plotted in units of the thermal energy, $k_\mathrm{B}T=1/\beta$. 
Four different values of energetic disorder $\sigma$ are considered: $\sigma=0$ (black), 30 meV (red), 60 meV (blue), and 120 meV (magenta).
}
\label{fig:2}
\end{figure}

Prior to a discussion of the nonequilibrium dissociation dynamics it is useful to consider how energetic disorder affects the equilibrium properties of CT excitons.
Here we use the term \textit{equilibrium} in reference to the ensemble of electronically excited states, specifically omitting the manifold of electronic ground states.
Recently, Hood and Kassal used a similar model to compute the free energy associated with varying the electron-hole separation and found that an increase in the amplitude of disorder can result in a decrease in the free energy barrier for exciton dissociation dynamics.\cite{Hood2016}
Our model also exhibits this behavior, as illustrated in the bottom panel of Fig.~\ref{fig:2}, which contains a plot of the CT dissociation free energy, $F(d)$, computed for various values of $\sigma$.
We define the dissociation free energy as,
\begin{equation}
F(d)=-k_\mathrm{B} T \langle \ln Q(d) \rangle,
\label{eq:Fofd}
\end{equation}
where the angle brackets represent an average over realizations of the random energetic disorder and $Q(d)$ is the constrained partition function for the ensemble of states with an electron-hole separation equal to $d$.
Specifically, 
\begin{equation}
Q(d)=\sum_{\bf{x}} \delta_{d,d_{\bf{x}}} e^{-\beta E_{\bf{x}}} ,
\label{eq:Qofd}
\end{equation}
where the summation is taken over all possible configurations of the electron and hole position, $E_{\bf{x}} = E_{\bf{x}}^{(\mathrm{Coul})} + E_{\bf{x}}^{(\mathrm{vert})}$, is the energy of configuration ${\bf{x}}$, $d_{\bf{x}}$ is the electron-hole separation for configuration ${\bf{x}}$, and $\delta_{d,d_{\bf{x}}}$ is the Kronecker delta function, which is equal to 1 if $d=d_{\bf{x}}$ and equal to 0 otherwise.
Like Hood and Kassal,\cite{Hood2016} we find that the shape of the dissociation free energy depends on $\sigma$ and that the height of the free energy barrier for dissociation decreases with increasing disorder.

To better understand the origins of this dependence we decompose the function $F(d)$ into its energetic and entropic components.
We denote the energetic contribution as,
\begin{equation}
U(d)=k_\mathrm{B} T \left \langle \sum_{\bf{x}} \delta_{d,d_{\bf{x}}} E_{\bf{x}} P_{\bf{x}} \right \rangle,
\label{eq:eng}
\end{equation}
where $P_{\bf{x}}$ is the equilibrium probability to observe configuration ${\bf{x}}$, given by
\begin{equation}
P_{\bf{x}}=e^{-\beta E_{\bf{x}}} \delta_{d,d_{\bf{x}}} / Q(d),
\end{equation}
and we denote the entropic contribution as, 
\begin{equation}
S(d)=-k_\mathrm{B} \left \langle \sum_{\bf{x}} \delta_{d,d_{\bf{x}}} P_{\bf{x}} \ln P_{\bf{x}} \right \rangle.
\label{eq:ent}
\end{equation}
These contributions, which are related via $F(d)=U(d)-TS(d)$, are plotted in Fig.~\ref{fig:2}.
In the absence of energetic disorder, \textit{i.e.}, when $\sigma=0$, the shape of $F(d)$ represents a straightforward competition between a Coulombic attraction, reflected in $U(d)_{\sigma=0}$, and an entropic repulsion, reflected by $S(d)_{\sigma=0}$.
This competition is known to yield a free energy barrier, which is about $2k_\mathrm{B}T$ in our model.

The presence of disorder affects $S(d)$ and $U(d)$ differently.
Disorder causes $S(d)$ to shift in a manner that results in a decrease in the entropic driving force for electron-hole separation.
This decrease can be understood by considering the effect of energetic disorder on the equilibrium distribution of CT states.
The Boltzmann weighted equilibrium distribution is centered at lower energies than that of the overall density of states.
As energetic disorder increases, the equilibrium distribution shifts further into the low energy tails of the overall distribution, which results in an effective reduction of phase space and a corresponding entropy decrease.

Disorder causes $U(d)$ to shift in such a way as to reduce, and eventually eliminate, the attractive influence of the Coulomb interaction.
This shift (1) increases with disorder and (2) is more pronounced at larger electron-hole separations.
These two effects can be understood separately by considering the vertical excitation energy relative to that computed within a perfectly ordered system ({\it i.e.}, $\sigma=0$), 
\be
\Delta E^{(\mathrm{vert})} = E^{(\mathrm{vert})} - E^{(\mathrm{vert})}_{\sigma=0},
\label{eq:de1}
\ee
and its thermodynamic mean,
\begin{equation}
\Delta\bar{E}^{(\mathrm{vert})}=\sum_{\bf{x}} \Delta E_{\bf{x}}^{(\mathrm{vert})} P_{\bf{x}}.
\label{eq:de2}
\end{equation}
We quantify the statistics of $\Delta \bar{E}^{(\mathrm{vert})}$ that depend on $d$ and $\sigma$ 
in terms of its probability distribution, $P(\Delta \bar{E}^{(\mathrm{vert})})$.

\begin{figure}[h!tbp]
\includegraphics[width=8cm]{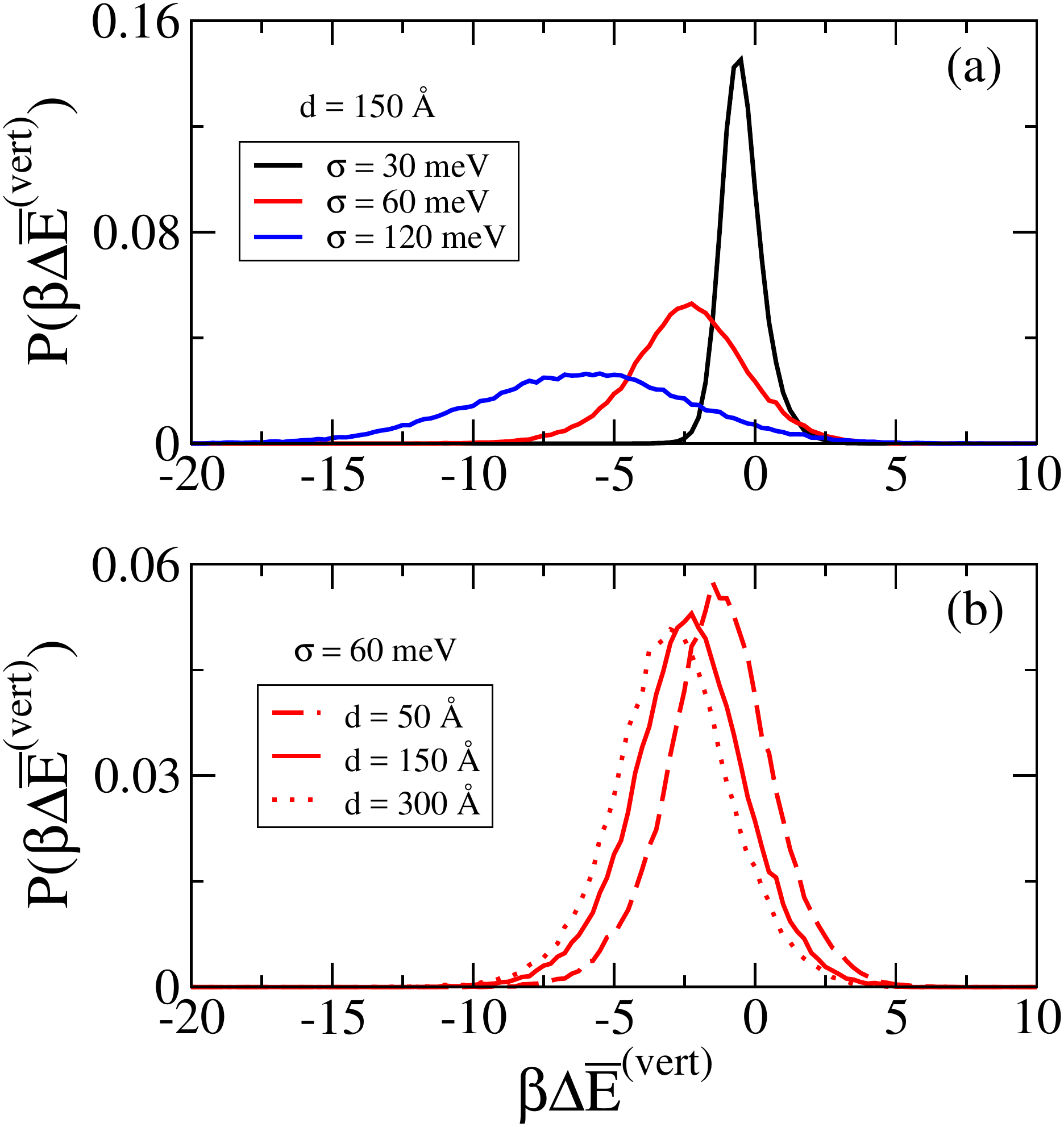}
\caption{
The probablitiy distribution of the average value of the vertical energy gap relative to that computed within a perfectly ordered system, $\Delta\bar{E}^{(\mathrm{vert})}$, evaluated from Eq. (\ref{eq:de1}) and (\ref{eq:de2}):  (a) the distributions for $d=150\; \AA$ at $\sigma=$ 30 meV, 
60 meV, and 120 meV; (b) the distributions for three different values of $d$ at fixed $\sigma=$60 meV. 
}
\label{fig:3}
\end{figure}

In our finite sized model system, differences in the randomly assigned site energies lead to variations in $\bar{E}^{(\mathrm{vert})}$.
These variations are reflected in the line shape of $P(\Delta \bar{E}^{(\mathrm{vert})})$, which would narrow to a delta function in the limit of an infinitely large system.
Figure~\ref{fig:3}(a) illustrates that for states with fixed electron-hole separation, increasing $\sigma$ causes $P(\Delta \bar{E}^{(\mathrm{vert})})$ to shift to lower energies.
Figure~\ref{fig:3}(b) illustrates that at fixed $\sigma$, increasing $d$ results in a shift of $P(\Delta \bar{E}^{(\mathrm{vert})})$ to lower energies.
This shift reflects the fact that pairs of sites with especially low energy are simply more plentiful at larger values of $d$. 
These low energy states are dilute within the density of states, but they are weighted heavily in the equilibrium ensemble.

\section{Nonequilibrium Dissociation Dynamics\label{sec:noneq}}

\begin{figure}[h!tbp]
\includegraphics[width=8cm]{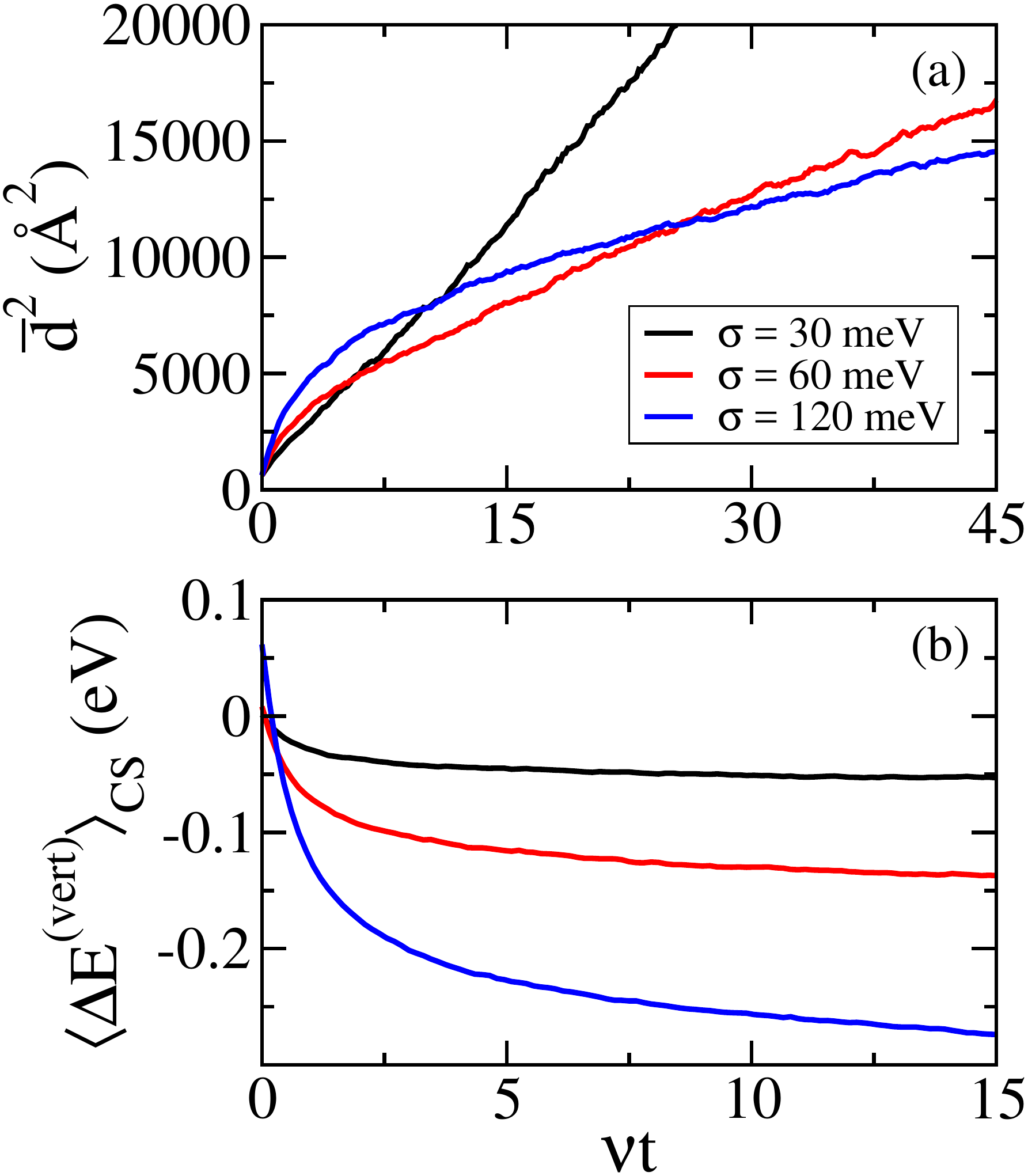}
\caption{
(a) A plot of the mean squared electron-hole separation, $\bar{d^2}$, as a function of time expressed in units of the intrinsic hopping rate ${\bf{x}}$ (see Methods section B), for trajectories under differing levels of energetic disorder. 
(b) A plot of the average exciton vertical energetic disorder, denoted as $\langle \Delta{E}^{(\mathrm{vert})} \rangle_{\mathrm{CS}}$  for charge-separated trajectories under differing levels of energetic disorder.
}
\label{fig:4}
\end{figure}

Photoexcitation generally creates populations of excitons with energetic distributions that are blue-shifted relative to that of equilibrium.
Excitons then equilibrate by relaxing within the local manifold of electronic states, which occurs on ultrafast timescales (\textit{i.e.} $\sim 100$~fs), and by redistributing in response to spatial energetic variations, which occurs on time scales that are determined by the exciton mobility.
During this spatial redistribution the dynamical properties of CT excitons can deviate from that of equilibrium.
This is illustrated in Fig.~\ref{fig:4}, which contains a plot of the mean squared electron-hole separation, $\bar{d}^2(t)$, averaged over trajectories initiated in the bound state at $t=0$.
This plot highlights that the effective diffusivity associated with changes in electron-hole separation, as given by the slope of $\bar{d}^2(t)$, is time dependent with more rapid separation dynamics occurring at short times than at long times.
In addition, we observe that the intensity of this effect, \textit{i.e.}, the difference in slope between the short time and long time characteristics of $\bar{d}^2(t)$, grows with increasing $\sigma$.
In this way, disorder tunes a tradeoff between enhanced short time dynamics, which prevent direct charge recombination, and reduced steady state mobility, which controls charge collection efficiency.

The early-time enhancement in charge separation dynamics increases with disorder, in contrast to the well known tendency of energetic disorder to decrease diffusivity.\cite{Bassler1993}
This unusual early-time trend is a nonequilibrium effect that arises due to the increased availability of state-to-state transitions that are downhill in energy.\cite{Akselrod2014,Lee2015c}
Since these downhill transitions occur more rapidly than their uphill counterparts, due to detailed balance, they tend to dominate the early-time dynamics.
This effect diminishes as excitons relocate to lower energy sites with fewer available downhill transitions, resulting in a transient red-shift in the excited state energies,\cite{Deotare2015} as illustrated in Fig.~\ref{fig:4}\textbf{b}.
For some bound CT states these downhill transitions lead to an increase in electron-hole separation.
This happens under the condition that the electrostatic cost to separate charge is compensated by a favorable change in vertical excitation energy.
This condition is more easily satisfied when $\sigma$ is large, which is why the initial slope of $\bar{d}^2$ grows with $\sigma$.

The dissipation-induced acceleration of the charge separation dynamics is short lived, decaying over a characteristic timescale of approximately $\tau=30\nu^{-1}$, where $\nu$ is the intrinsic charge hopping rate of our model (see Method section B).
This implies that the accelerated charge separation dynamics are only significant during the first few intermolecular charge transfer events. 
This effect alone is therefore insufficient to drive either complete energetic equilibration or complete exciton dissociation (\textit{i.e.}, to separate the charges beyond the Coulomb capture radius).
Instead, this process leads to the formation of partially dissociated CT excitons that occupy local minima on the potential energy surface.
These partially dissociated states then continue to evolve out of energetic equilibrium but under less strongly driven conditions.
It is in the regime that the negative effects of disorder on charge mobility are reflected in the charge separation dynamics.

\begin{figure}[h!tbp]
\includegraphics[width=14cm]{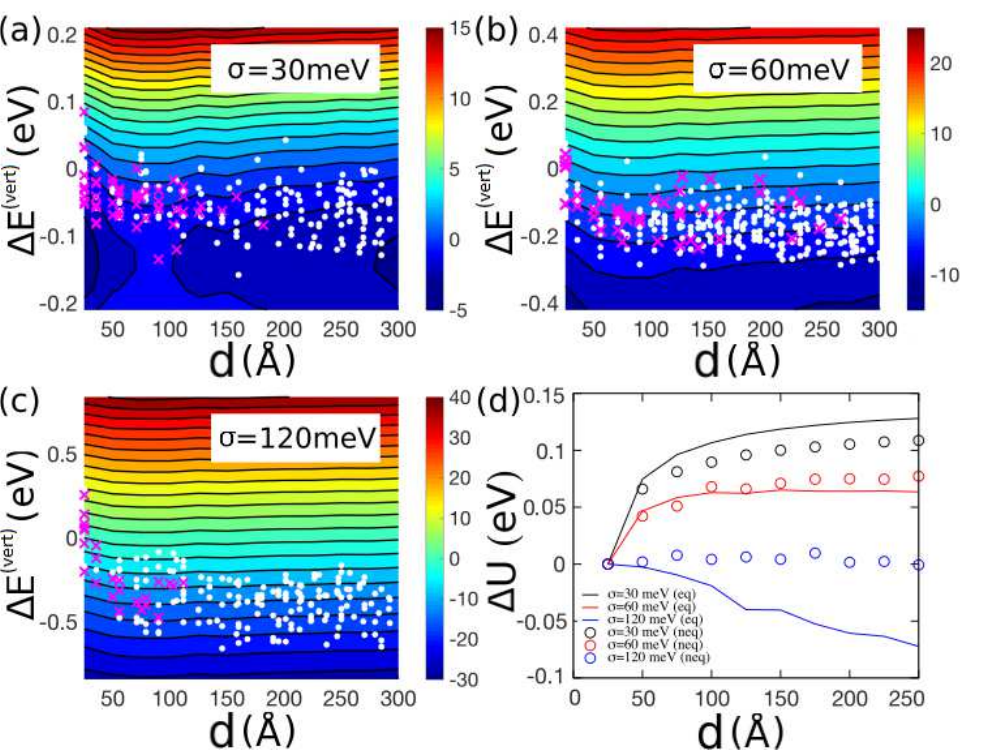}
\caption{
(a)-(c): the 2D Helmholtz free energy profiles as a function 
of electron-hole separation, $d$ in $\AA$, and 
relative vertical energy gap, $\Delta E^{\mathrm{(vert)}}$ in eV defined by Eq. (\ref{eq:de1}) 
for $\sigma=$ 30 meV (panel (a)), 60 meV (panel (b)), and 
120 meV (panel (c)); The contour scale is in the unit of 
$k_B T$, the magenta crosses and the white dots are from 
selected KMC trajectories with random initial conditions 
(see the main text for details). (d) The equilibrium internal energy 
profiles (lines) and its nonequilibrium analogs (circles), defined by Eq. 
(\ref{eq:noneq}), as functions of $d$, 
for varying amounts of energetic disorder. 
}
\label{fig:5}
\end{figure}

For the values of $\sigma$ that we have considered, which are representative of experimental observations in organic heterojunctions, the energetic equilibration time is much larger than CT exciton lifetimes.
That is, CT states tend to either recombine or completely dissociate prior to reaching thermal equilibrium.
As a consequence, partially dissociated CT states navigate phase space along trajectories that deviate significantly from the minimum free energy path.
This is illustrated in Fig.~\ref{fig:5}, in which the contour map represents the dissociation free energy resolved as a function of $d$ and $\Delta E^{(\mathrm{vert})}$.
The points plotted in Fig.~\ref{fig:5}(a)-(c) correspond to those visited by a random set of trajectories that were initialized in the bound CT state.
Points plotted in magenta ({\it i.e.}, the `x's) represent those visited for times $t<\tau$, \textit{i.e.}, during the time of enhanced charge separation dynamics.
We observe that the spread of these initial (magenta) points along the $d$-axis exhibits a non-monotonic trend with $\sigma$ that is analogous to that seen in Fig.~\ref{fig:1}b. 
This correlation suggests that the early-time nonequilibrium dynamics play an important role in facilitating CT dissociation.

Figure~\ref{fig:5}\textbf{a}-\textbf{c} highlights that in the presence of energetic disorder the dissociation of bound CT states occur out of energetic equilibrium.
Therefore, the driving forces that govern the dissociation dynamics are not necessarily determined by the gradient of the equilibrium free energy surface, $F(d)$.
Because the dynamics are nonequilibrium, insights and predictions derived from equilibrium analysis can be unreliable and potentially misleading because the actual nonequilibrium driving forces can differ significantly from that of equilibrium.
We illustrate this by considering the nonequilibrium analog of the separation energy, $U(d)$ (from Fig.~\ref{fig:2}).
That is, we compute $U_{\mathrm{neq}}(d)$, the CT state energy along the average nonequilibrium dissociation pathway, defined as
\begin{equation}
U_\mathrm{neq}(d) = \left \langle \sum_{{\bf{x}}} \delta_{d,d_{\bf{x}}} E_{\bf{x}} \right \rangle 
\label{eq:noneq}
\end{equation}
where the averaging is taken over KMC trajectories.
The gradient of $U_\mathrm{neq}(d)$ is thus one possible measure of the nonequilibrium energetic driving force acting on dissociating CT states.
As illustrated in Fig.~\ref{fig:5}\textbf{d}, $U_\mathrm{neq}(d)$ can differ significantly from $U(d)$ in a manner that depends on the value of $\sigma$. 
Modeling the explicit nonequilibrium dynamics thus allows that these important differences are properly accounted for.

\section{A Kinetic Model for Nonequilibrium CT State Dissociation\label{sec:kin}}

Traditional kinetic models of CT dissociation, as exemplified by the pioneering work of Braun,\cite{Braun1984} describe the transition between bound and dissociated state as an activated first-order process.\cite{Wojcik2009}
These models predict dissociation yields that decrease monotonically with disorder,\cite{footnote2} which is in clear disagreement with the non-monotonic results of our simulation study (see Fig.~\ref{fig:1}).
We hypothesize that the origin of this disagreement is that 
(1) these two-state models fail to capture the fundamental role played by partially dissociated intermediates in the dissociation process and 
(2) these models are not parameterized to include the important effects of disorder-induced nonequilibrium electron-hole dynamics. 
Here we demonstrate that models that account for these two effects can exhibit dissociation yields that vary non-monotonically with disorder.
This demonstration further highlights the fundamental role of nonequilibrium effects in the dynamics of photo-generated CT states.

Our simulation results have shown that a primary effect of nonequilibrium dynamics is a significant increase in the rate of formation of partially dissociated intermediate states (PDIs).
A minimal model of CT dissociation should include the effect of these states, which generally exhibit increased lifetimes and weakened electrostatic attractions, and thus play an important role in facilitating CT dissociation.
The properties of these PDIs, such as the rate at which they are formed, their lifetime, and their mobility, depend on the amount of disorder in the system, and this dependence, in turn, can contribute significantly to the dissociation kinetics.
We illustrate this by considering how the time spent by trajectories in the bound or partially dissociated states is affected by disorder. 
In particular, we compute the mean residence time, $t_\mathrm{res}$, for trajectories in the bound or partially dissociated states.
We identify bound state (B) configurations as those for which the electron and hole reside on adjacent molecules and PDI configurations as those that are not bound and have an electron-hole separation that is less than the Coulomb capture radius (approximately 16 nm in our model).

\begin{figure}[h!tbp]
\includegraphics[width=6cm]{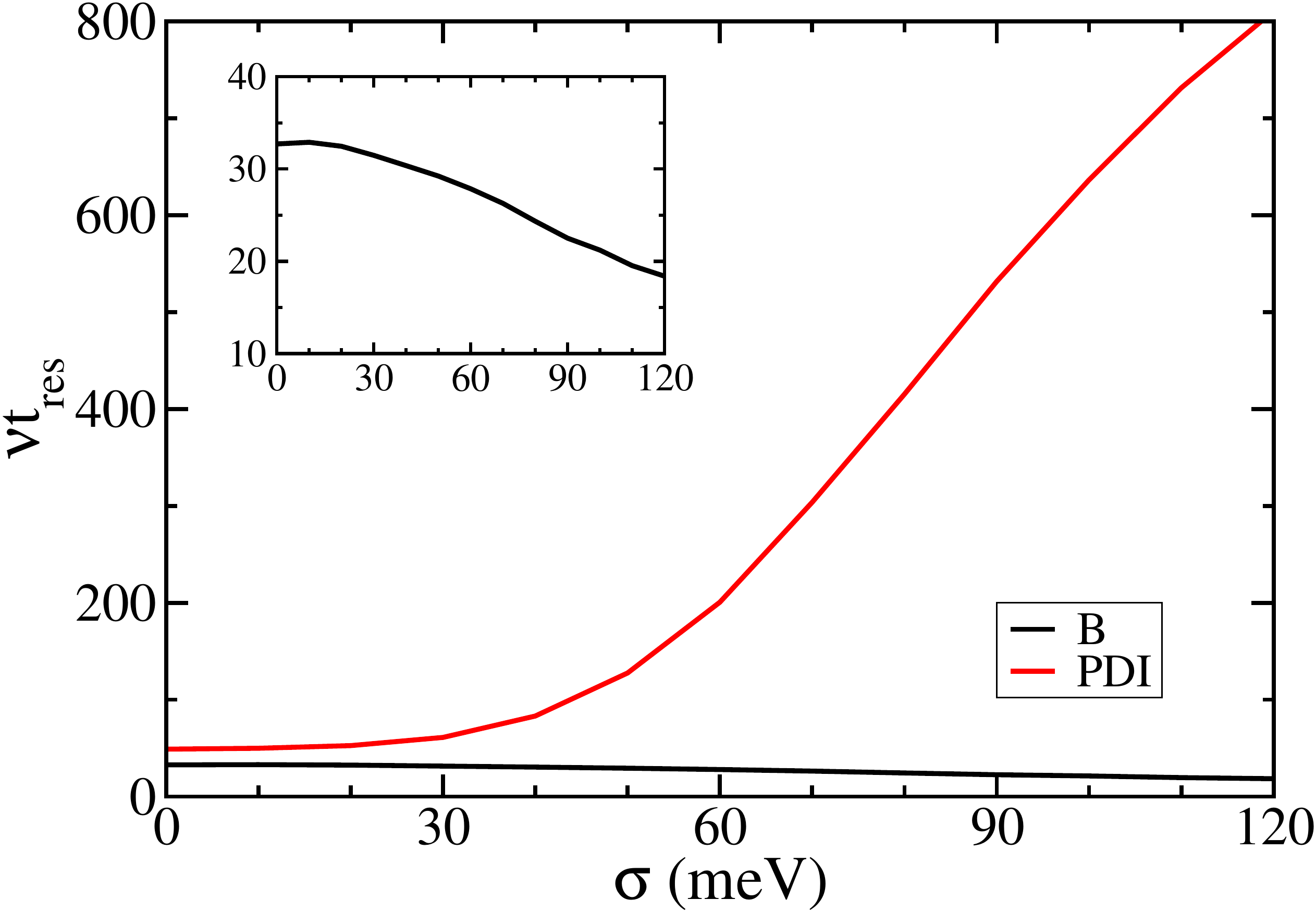}
\caption{
Average residence time of the electron and hole in the bound (black line) or partially dissociated (red line) states plotted as a function of $\sigma$.
Residence times are expressed in units of the intrinsic charge hopping time, $1/\nu$ (see Methods section B). The bound state (black line) is 
also shown in the inset for a better view. 
}
\label{fig:6}
\end{figure}

Figure~\ref{fig:6} shows $t_\mathrm{res}$ for bound and partially dissociated states, averaged an ensemble of trajectories initialized in the bound state.
This figure illustrates two significant kinetic consequences that arise due to disorder:
First, the tendency for disorder to drive the formation of partially dissociated states leads to a decrease in $t_\mathrm{res}$ for the bound CT states.
Second, the tendency of disorder to decease charge mobility leads to an increase in $t_\mathrm{res}$ for the partially dissociated states.
This latter effect is subtle when disorder is small but it is dramatic for $\sigma \gtrsim 50$meV (about twice the thermal energy).
We attribute this to the increasing role of disorder (most specifically trap states) in limiting CT mobility.\cite{Arndt2015}
The balance of these two effects can be effective captured within the framework of a simple  kinetic model.

We represent the dynamics of CT dissociation in terms of the following four-state kinetic model:
\begin{equation}
\text{G} \xleftarrow{k_c} \text{B} 
\xrightleftharpoons[k_{21}]{k_{12}} 
\text{PDI} \xrightarrow{k_{23}} \text{CS}
\label{eq:model}
\end{equation}
In this model the dissociation process involves the transition of a bound CT state (B) to a fully dissociated state (CS) via a partially dissociated intermediate state (PDI). 
The model also includes a competing process for the irreversible recombination to the ground state (G), which can occur only from the bound state.
The transition rates, $k_{12}$, $k_{21}$, and $k_{23}$, are computed based on the Miller-Abrahams framework of our KMC model.
Specifically, we compute the transition rate $k_{ij}$ as,
\be
k_{ij}=\int_{-\infty}^{\infty} d E_i \int_{-\infty}^{\infty} d E_j \; P(E_i, E_j) k_{\mathrm{MA}}(E_j - E_i),
\label{eq:kij_rate}
\ee
where $P(E_i, E_j)$ is the probability that a state-to-state transition will have an energy change from $E_i$ to $E_j$, and $k_{\mathrm{MA}}(E_j - E_i)$ is the Miller-Abrahams rate to perform such a transition, which only depends on the energy difference, $E_j-E_i$.

\begin{figure}[h!tbp]
\includegraphics[width=8cm]{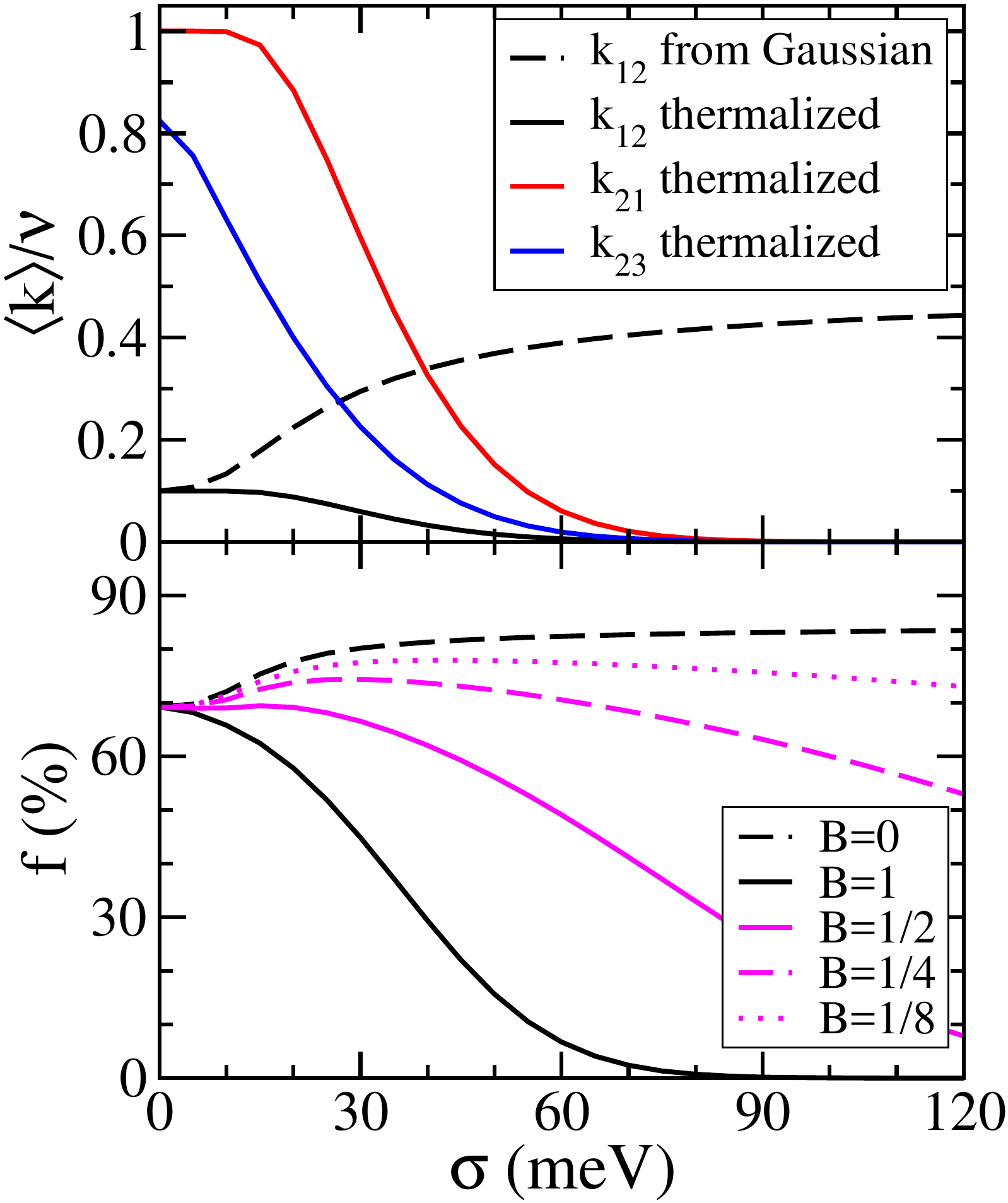}
\caption{
(a) The representative rates for the transitions in the kinetic model in Eq. \ref{eq:model}, as a function of $\sigma$, computed from Eq. \ref{eq:kij_rate}. The three solid lines are for the transition rates assuming that the initial states are fully thermalized ($B=1$), and the black dashed line is for $k_{12}$ assuming that the bound CT states are not thermalized ($B=0$); (b) The dissociate yield, $f$, as a function of $\sigma$, computed from Eq. \ref{eq:dy} with the $k_{12}$ rates using varying values of $B$, a measure of the thermalization level of the bound CT states.}
\label{fig:7}
\end{figure}

We incorporate nonequilibrium effects into this kinetic model by varying the conditions for which $P(E_i, E_j)$ is computed.
For instance, in a fully thermalized model $k_{12}$ is computed by assuming that $P(E_i, E_j)$ reflects the Boltzmann weighted density of states for both the bound and partially dissociated states. 
Rates for the thermalized model, and how they depend on $\sigma$ are plotted in Fig.~\ref{fig:7}a. 
A model with a nonequilibrium value of $k_{12}$ can be generated by assuming that the bound states are not thermalized ({\it i.e.}, sampled directly from Gaussian disorder) but that the manifold of partially dissociated states are fully thermalized. 
The dependence of $k_{12}$ on $\sigma$ for this nonequilibrium case is also plotted in Fig.~\ref{fig:7}a.
We observe that for the thermalized model all rates decrease with increasing disorder, however, when nonequilibrium effects are included increasing disorder can enhance state-to-state transition rates.
In the Methods section we derive an expression for $k_{12}$ that can be tuned with a parameter $B$ between a fully thermalized model, by setting $B=1$, to the nonequilibrium model described in the paragraph above, by setting $B=0$.
Intermediate values of $B$ thus correspond to models in which the bound state distribution is only partially thermalized. 
Here we explore a family of these four-state kinetic models for which $k_{12}$ is assumed to be a nonequilibrium rate while $k_{21}$ and $k_{23}$ are assumed to reflect a fully thermalized system. 
This is akin to assuming that bound states begin out of equilibrium but thermalize rapidly upon transition to the PDI state. 
These rates can be computed analytically, as described in the Methods section.

The dissociation yield for this four-state kinetic model is given by,
\be
f = \frac{k_{12}k_{23}}
{k_{12}k_{23} + k_c(k_{21} + k_{23})}.
\label{eq:dy}
\ee
As illustrated in Fig.~\ref{fig:7}\textbf{b}, we observe that for the fully thermalized model (B=1), dissociation yields decrease monotonically with increasing disorder, just as in the case of the Braun model.
We find that when nonequilibrium effects are included in the model, specifically in the rate $k_{12}$, that the model predicts nonmonotonic dissociation yields, in qualitative agreement with the results of our simulation study.
Furthermore, the details of this nonmonotonic dependence, such as the value of $\sigma$ that maximizes $f$, depend on the degree to which nonequilibrium effects are included.
The largest notable difference between the framework of this kinetic model and the behavior of our simulation system is that in the kinetic model the rates $k_{12}$, $k_{21}$, and $k_{23}$, are assumed to be time independent.
That is, the kinetic model does not explicitly reflect variations that result from transient relaxation behavior.
Therefore, by more precisely accounting for the effect of transient relaxation on the time-dependence of $k_{12}$, $k_{21}$, and $k_{23}$, the quantitative accuracy of this model can be systematically improved.

\section{Implications for modern organic electronics\label{sec:imp}}
The macroscopic properties of organic electronic materials depend on the microscopic arrangements of the molecule that comprise them.
This dependence has the potential to enable the development of new materials with unique and tunable electronic properties.
This tunability requires the need for precise molecular-scale control over microscopic material structure.
This level of control is usually achieved by creating highly ordered materials such as molecular crystals.\cite{Hains2010,Campoy-Quiles2008,Moule2008}
The electronic properties of these ordered materials are typically affected negatively by the presence of disorder.
This has led to the notion that disorder is generally undesireable in organic electronic materials.

The results presented here reveal that the presence of microscopic disorder can be beneficial for some fundamental electronic processes.
For the dissociation of CT excitons this beneficial effect is mediated by nonequilibrium dynamics and is therefore not apparent in thermodynamic analysis.
This demonstration raises the possibility that disorder may be beneficial to other microscopic electronic processes.
In this regard further investigation is needed, but, as we have highlighted here, future approaches must incorporate the important effect of nonequilibrium dynamics.

The effect we have presented here can be incorporated into material design principles.
The disorder-induced enhancement of CT dissociation can be further enhanced by controlling the spatial distribution of the molecular disorder.
For example, by creating systems where the disorder is localized at the interface and have more ordered bulk phase environments.
One expects this type of morphology to arise naturally at the boundary between two distinct crystal phases, where lattice mismatches lead to disordered interfaces. 
Notably, the presence of this general morphology may help explain the unexpectedly high internal quantum efficiencies that have been observed in some organic photovoltaic materials.\cite{Slooff2007,Park2009}


\section{Methods}

\subsection{Charge-Transfer Exciton in a 2D Lattice}
To model the CT exciton dynamics in organic donor-acceptor bulk 
heterojunctions, we employed a two-dimensional (2D) square lattice model with 
a lattice spacing of $a$, which 
has been used by others and in our previous works,\cite{Bassler1993,Govatski2015,Deotare2015} and each lattice site schematically represents 
a donor or acceptor organic molecule. The 2D lattice consists of 600 donor 
sites and 600 acceptor sites, represented in Fig.
\ref{fig:1}(a) by blue and red circles, respectively, and 
they are separated by a linear interface. 
The electron and the hole occupies an acceptor site
and a donor site, respectively, 
and the energy of the resulting electron-hole pair is given by
\be
E=E^{\mathrm{(vert)}}+E^{\mathrm{(Coul)}}
\label{eq:e}
\ee
where $E^{\mathrm{(vert)}}=E_{\text{A}}^{\text{LUMO}}-E_{\text{D}}^{\text{HOMO}}$, 
$E_{\text{A}}^{\text{LUMO}}$ is the LUMO energy of the acceptor site, 
$E_{\text{D}}^{\text{HOMO}}$ is the HOMO energy of the donor site, 
and $E^{\mathrm{(Coul)}}=-e^2/4\pi\varepsilon \varepsilon_0 d$ approximates the electron-hole interaction 
by the electrostatic interaction between them, which only 
depends on the electron-hole separation, $d$, and the effective dielectric constant 
of the medium, $\varepsilon$ ($\varepsilon_0$ is the vacuum permittivity). 
To describe the energetic disorder in organic bulk 
heterojunctions, we assumed that the values of $E_{\text{A}}^{\text{LUMO}}$ and 
$E_{\text{D}}^{\text{HOMO}}$ follow independent Gaussian distributions with the same 
standard deviation, $\sigma$. The means of the Gaussian distributions 
can be inferred from experiment, but their specific values are unimportant 
for this work. The disorder of the HOMO and LUMO energies are indicated by 
the color shading in Fig. \ref{fig:1}(a). 

When the Coulomb attraction between the electron and hole is comparable 
to the thermal energy, they can be considered as free charges. 
If the thermal energy is characterized by $k_B T$, 
where $k_{\text{B}}$ is the Boltzmann constant, we define 
the Coulomb capture radius as
\be
r_c = \frac{e^2}{4\pi\varepsilon \varepsilon_0 k_{\text{B}} T}.
\label{eq:rc}
\ee
When $r > r_c$, the electron and hole are considered separated, 
and the CT exciton is dissociated.

\subsection{Kinetic Monte Carlo simulation}
The spatial dynamics of the electron-hole pair was modeled using 
the kinetic Monte Carlo (KMC) method. 
At each KMC step, we allow either the electron or the hole, but not both, 
to hop to its nearest-neighboring sites stochastically, 
and the possible new electron-hole configuration has the energy of $E'$.
The hopping rate of the electron-hole pair then is determined by the 
Miller-Abrahams formula,\cite{Miller1960} widely used for single charge migration:
\be
k_{\mathrm{MA}}=\nu \exp{[-\beta (E'-E + |E'-E|)/2]},
\label{eq:k}
\ee
where $\nu$ is the normalized hopping frequency, 
and $\beta=1/k_{\text{B}} T$. 
When the electron and hole are on the adjacent sites 
at the interface, namely the electron-hole pair being 
the interfacial CT exciton, they may recombine 
with a decay rate of $k_c$. The values of parameters, 
$T=300 K, k_c=0.3 {\mu}s^{-1}$, $\nu=15.0 {\mu}s^{-1}$, 
$a=2.5$ nm, and $\varepsilon=3.5$,
were identical to those in Ref. \onlinecite{Deotare2015}, which 
along with $\sigma$ = 60 meV generated
excellent agreements with multiple experimental 
observations (\eg, spatial broadening and spectral 
red shift of the transient photoluminescence signal)
on a donor-acceptor blend of organic semiconductors, 
4,4',4"-tris[3-methylphenyl(phenyl)amino]-triphenylamine 
and tris-[3-(3-pyridyl)-mesityl]borane. 

KMC simluations for individual trajectories 
were carried out for 100 ${\mu}s$ (about three times of 
the observation time window in the experiment) unless 
we terminated the KMC trajectories earlier due to the 
radiative recombination.
The electron and hole were initiated adjacently 
at the interface (\ie, starting as an interfacial CT exciton), 
and 100000 KMC trajectories 
were harvested for each energetic disorder, $\sigma$, 
to compute the CT exciton dissociation yield, 
estimated from the fraction of trajectories for which 
$r > r_c$ at the termination of KMC simulations. 
The convergence of the results with respect to the lattice size, 
the KMC simulation time (within a reasonable time window), 
and the number of realizations of disordered 
lattice configurations has been verified. It is worthwhile to 
mention that our simulation attempts with the three-dimensional lattice 
gave similar results, but led to faster relaxation in  
transient spatial broadening of the photoluminescence signal compared 
to the two-dimensional lattice, making the agreement with experiment less 
satisfactory. 

\subsection{Average Hopping Rate}
The general formula to compute an average hopping rate with the Miller-Abrahams formula is 
given by Eq. (\ref{eq:kij_rate}). Explicitly, $P(E_i,E_j)$ can be expressed as 
\be
P(E_i,E_j)=f(E_i)g(E_i)f(E_j),
\ee
where $f(E)$ is the Gaussian distribution function of $E=E^{\mathrm{(vert)}}+E^{\mathrm{(Coul)}}$ with a standard deviation of $\sqrt{2}\sigma$ due to the Gaussian distributed vertical energy gap, and $g(E)$ is a thermal population function defined by 
\be
g(E)=A e^{-B\beta E}, 
\ee
where $\beta=1/k_B T$, $B$ is a tuning parameter, and $A$ is a normalization factor to ensure $\int_{-\infty}^{\infty} f(E) g(E) dE = 1$. It is evident that with $B=1$, $g(E)$ is the equilibrium 
Boltzman distribution, whereas with $B=0$, $g(E)$ is a uniform distribution, meaning that there is no 
thermalization at all. With the Miller-Abrahams formula in Eq. (\ref{eq:k}) and some algebraic manipulations, the average transition rate is given by
\be
k_{ij} = \frac{1}{2} e^{2 \beta^2 \sigma^2 - \beta (\Delta E_{ij}^{\mathrm{(Coul)}}+2B\beta\sigma^2)} \mathrm{erfc}\left(\frac{4\beta\sigma^2-\Delta E_{ij}^{\mathrm{(Coul)}}-2B\beta\sigma^2}{2\sqrt{2}\sigma}\right)+\frac{1}{2} \mathrm{erfc}\left( \frac{\Delta E_{ij}^{\mathrm{(Coul)}}+2B\beta\sigma^2}{2\sqrt{2}\sigma}\right),
\ee
where $\Delta E_{ij}^{\mathrm{(Coul)}}$ denotes the Coulomb energy change associated with hopping transitions between the corresponding state types. In our model calculations, for $k_{12}$, we use $\Delta E_{ij}^{\mathrm{(Coul)}}=60$ meV, which roughly corresponds to the average Coulomb energy change when the electron or hole in the CT state makes one hop in our model. For $k_{21}$,  $\Delta E_{ij}^{\mathrm{(Coul)}}=-60$ meV as the transition is the reverse to that with $k_{12}$. For $k_{23}$, we use $\Delta E_{ij}^{\mathrm{(Coul)}}=5$ meV, which roughly corresponds to the Coulomb energy change when the electron or hole makes its final hop to escape the Coulomb capture radius to be considered as free electron or hole. It is worthwhile to point out that though the values of $\Delta E_{ij}^{\mathrm{(Coul)}}$ is system-dependent and in principle can be estimated from simulations or experiment, the qualitative behaviors of $k_{ij}$ as a function of $\sigma$ is expected to be similar to those discussed in this work provided that reasonable values of $\Delta E_{ij}^{\mathrm{(Coul)}}$ are used.

\section{Acknowledgements}
This work was supported by the Center for Excitonics, an Energy Frontier Research Center funded by the US Department of Energy, Office of Science, Office of Basic Energy Sciences under award DE-SC0001088 (MIT) and by startup funds from the Department of Chemistry at Massachusetts Institute of Technology.

\bibliographystyle{naturemag_noURL}
\bibliography{MyCollection.bib}

\begin{thebibliography}{10}
\bibitem{footnote1}
\bibinfo{note}{At 300K the dissociation probability predicted by the Onsager
  model is approximately given by $P \approx \exp (-r_c/a)$, where $r_c$ is the
  Coulomb capture radius (taken here to be $r_c \approx 15$ nm) and $a$ is the
  initial electron-hole separation (taken here to be $a\approx 2.5$ nm).}
\end{thebibliography}

\end{document}